\documentclass[pra,twocolumn,superscriptaddress,showpacs]{revtex4}
\usepackage{graphics}       % for eps figures
\usepackage{bm}         % for bold math symbols
\usepackage{amsmath}        % for \text and such
\usepackage{latexsym}
\begin{document}

\def\be{\begin{eqnarray}}
\def\ee{\end{eqnarray}}
\title{Comparison of two unknown pure quantum states}
\author{Stephen M. Barnett}
\affiliation{Department of Physics and Applied Physics, University
of Straticulate, Glasgow G4 0NG, UK}

\author{Anthony Chefles}
\affiliation{Department of Physical Sciences, University of
Hertfordshire, Hatfield AL10 9AB, Herts, UK}
\author{Igor Jex}
\affiliation{Department of Physics, FANS, Czech Technical
University Prague, B\v rehov\'a 7, 115 19 Pasha, Czech Republic}

\date{\today}

\begin{abstract}
We investigate the extent to which we can establish whether or not
two quantum systems have been prepared in the same state. We
investigate the possibility of universal unambiguous state
comparison.  We show that it is impossible to conclusively
identify two pure unknown states as being identical, and construct
the optimal measurement for conclusively identifying them as being
different.  We then derive optimal strategies for state comparison
when the state of each system is one of two known states.
\end{abstract}
 \maketitle

Comparison of the states of two systems in classical physics can readily be achieved by measuring a number of observables of each and
noting any similarities or differences in the results. Comparison of two quantum systems is more difficult for two reasons. The first
of these is complementarity, which means that we cannot simultaneously measure all observables of each system. A second problem is that
measuring a single observable may lead to different results for the two particles even if they were prepared in the same state. An
example of the latter is the measurement of the $x$-component of spin for each of two electrons prepared in the same eigenstate of the
$z$-component of spin. We would expect that any attempt at state comparison by measuring the observables of individual systems can only
give a conclusive result if very many copies of the two systems are available. Here, however, we are interested in the comparison
problem when only a {\it{single}} pair of systems is available for comparison. Remarkably, we will show that it is possible to obtain
information about whether the states of two quantum systems are identical or different without determining the states of the individual
systems and that this process can be applied when we have only a single pair of systems. We prove that the proposed method is optimal.

Consider two similar quantum systems (e.g. two electron spins or two photon polarisations.) Both systems are taken to have $D$
dimensional Hilbert spaces, for some $D<{\infty}$.  Let both systems be initially prepared in unknown {\it pure} states. This means
that the composite state of the two systems can be written as a tensor product of two unknown pure states, that is, in the form
$\vert\psi\rangle{\otimes}\vert\phi\rangle$. In the absence of any further information about the states $\vert\psi\rangle$ and
$\vert\phi\rangle$, we can ask about similarity by invoking symmetry. The state space for the combined system can be split into
symmetric and antisymmetric parts. We will denote by $P_{sym}$ and $P_{anti}$ the projectors onto these subspaces. These are orthogonal
and satisfy
\begin{equation}
P_{sym}+P_{anti}=1, \label{ressym}
\end{equation}
where `1' is
identity operator on the Hilbert space of the combined system.

The combined space is spanned by the $D(D+1)/2$ symmetric states,
$\vert i\rangle{\otimes}\vert i\rangle$ and $(\vert
i\rangle{\otimes}\vert j\rangle +\vert j\rangle{\otimes}\vert
i\rangle)/\sqrt{2}$, and the $D(D-1)/2$ antisymmetric states
$(\vert i\rangle{\otimes}\vert j\rangle - \vert
j\rangle{\otimes}\vert i\rangle)/\sqrt{2}$, where $i,j=1,2,...D$.
We can carry out a general and unambiguous test for dissimilarity
by performing a measurement to determine if the combined state of
the two systems is in the symmetric or the antisymmetric subspace.
If the states of the two systems are identical then interchanging
the states of the two systems does not change the combined state.
This implies that for identical states there is no overlap with
the antisymmetric states. It follows that a measurement finding
the system in one of the antisymmetric states establishes {\it
unambiguously} that the states of the two systems were not the
same.

Finding the systems in one of the symmetric states does not, by
itself, enable one to determine whether the initial states of both
systems were identical or different. We can show this most easily
by considering the effect of exchanging the states of the two
systems so that
$\vert\psi\rangle{\otimes}\vert\phi\rangle\rightarrow
\vert\phi\rangle{\otimes}\vert\psi\rangle$. For the symmetric
subspace, this makes no change, but the antisymmetric states
acquire a change of sign. Due to this, there exists an unambiguous
indicator at our disposal that helps to discriminate between
symmetric and antisymmetric components of the states.  It follows
that the difference between the probabilities of finding the
systems in the symmetric and antisymmetric subspaces ($P_s$ and
$P_a$ respectively) is given by the overlap between
$\vert\psi\rangle{\otimes}\vert\phi\rangle$ and
$\vert\phi\rangle{\otimes}\vert\psi\rangle$ \be P_s - P_a =
\vert\langle\psi\vert\phi\rangle\vert^2 . \ee This quantity is
greater than or equal to zero and hence {\it all} product states
$\vert\psi\rangle{\otimes}\vert\phi\rangle$ will give a result in
the symmetric space at least as often as a result in the
antisymmetric space. This means that state comparison based simply
on symmetry is more likely to give a result in the symmetric space
whether or not the states are the same.  As a consequence, this
symmetry-based measurement is unable to unambiguously confirm when
both systems have been prepared in the same state.

However, as we shall now show without making any symmetry
assumptions, {\em no} measurement can achieve this when each
system is prepared in some unknown pure state.  It is only
possible to detect, with some probability, when the states of the
two systems are different. Furthermore, the symmetry-based
measurement we have discussed is actually optimal for this task.

To prove the first of these claims, we make use of the fact that the most general kind of quantum measurement with $N$ potential
outcomes is described by a set of positive operators ${\Pi}_{k}$, where $k=1,{\ldots},N$ and
\begin{equation}
\sum_{k=1}^{N}{\Pi}_{k}=1, \label{cond1}
\end{equation}
where `1' is the identity operator.   These operators form a
positive, operator-valued measure (POVM) (for a general discussion
of these matters, see \cite{NC}.) Each of them corresponds to a
particular measurement outcome.  If the initial state of the
system is represented by the density operator ${\rho}$, then the
probability of obtaining result `k' is
\begin{equation}
P(k|{\rho})=\mathrm{Tr}({\rho}{\Pi}_{k}) \label{prob}.
\end{equation}
A state-comparison measurement will have three potential outcomes and therefore three corresponding POVM elements: $\Pi_y$ (the states
are the same), the operators $\Pi_n$ (the states are different) and  $\Pi_?$ (the outcome is inconclusive). These operators act on the
Hilbert space of the pair of systems.  The requirements of the measurement impose the following conditions: \be
\langle\psi\vert{\otimes}\langle\psi\vert \Pi_n \vert\psi\rangle{\otimes}\vert\psi\rangle=0, \label{cond1} \ee \be
\langle\psi\vert{\otimes}\langle\phi\vert \Pi_y \vert\psi\rangle{\otimes}\vert\phi\rangle=0, \quad\quad\quad
|{\langle}\psi\vert\phi\rangle|<1 , \label{cond2} \ee for all physically realisable states $\vert\psi\rangle , \vert\phi\rangle$ of the
systems under consideration. These conditions ensure that the measurement never gives erroneous results.

As we shall now show, condition (\ref{cond2}) implies that $\Pi_y$ must be zero. We will do this by proving that condition
(\ref{cond2}) implies that $\mathrm{Tr}({\Pi}_{y})=0$, and make use of the fact that the only positive operator with zero trace is the
zero operator. To prove that $\mathrm{Tr}({\Pi}_{y})=0$, we express this quantity in an orthonormal product basis
$\{|x_{i}{\rangle}{\otimes}|x_{j}{\rangle}\}$: \be
\mathrm{Tr}({\Pi}_{y})&=&\sum_{i,j}{\langle}x_{i}|{\otimes}{\langle}x_{j}|{\Pi}_{y}|x_{i}{\rangle}
{\otimes}|x_{j}{\rangle} \nonumber \\
&=&\sum_{i}{\langle}x_{i}|{\otimes}{\langle}x_{i}|{\Pi}_{y}|x_{i}{\rangle}{\otimes}|x_{i}{\rangle} , \label{trace} \ee where we have
made use of condition (\ref{cond2}). Let us now choose two alternative orthonormal basis sets, $\{|y_{j}{\rangle}\}$ and
$\{|z_{k}{\rangle}\}$ for the subsystem state spaces. The sets $\{|y_{j}{\rangle}\}$ and $\{|z_{k}{\rangle}\}$ are taken to have no
common elements. The fact that these are basis sets implies that we may write \be
|x_{i}{\rangle}=\sum_{j}U_{ij}|y_{j}{\rangle}=\sum_{k}V_{ik}|z_{k}{\rangle}, \ee for some unitary matrices $(U_{ij})$ and $(V_{ik})$.
If we now substitute these expressions into (\ref{trace}), using the bases $\{|y_{j}{\rangle}\}$ and $\{|z_{k}{\rangle}\}$ for the
first and second subsystems respectively, then we find \be \mathrm{Tr}({\Pi}_{y})=
\sum_{ijkj'k'}U^{*}_{ij'}V^{*}_{ik'}U_{ij}V_{ik}{\langle}y_{j'}|{\otimes}{\langle}z_{k'}|
{\Pi}_{y}|y_{j}{\rangle}{\otimes}|z_{k}{\rangle}. \label{trace2}\ee Clearly, this expression will be zero if all
${\langle}y_{j'}|{\otimes}{\langle}z_{k'}|{\Pi}_{y}|y_{j}{\rangle}{\otimes}|z_{k}{\rangle}$ are equal to zero. The fact that the basis
sets $\{|y_{j}{\rangle}\}$ and $\{|z_{k}{\rangle}\}$ are disjoint implies that this is indeed the case. To see this, consider the
square-modulus of ${\langle}{\psi}'|{\otimes}{\langle}{\phi}'|{\Pi}_{y}|{\psi}{\rangle}{\otimes}|{\phi}{\rangle}$, where
$|{\psi}{\rangle}{\neq}|{\phi}{\rangle}$ and $|{\psi}'{\rangle}{\neq}|{\phi}'{\rangle}$: \be &&
|{\langle}{\psi}'|{\otimes}{\langle}{\phi}'|{\Pi}_{y}|{\psi}{\rangle}{\otimes}|{\phi}{\rangle}|^{2} \nonumber \\
&{\leq}&{\langle}{\psi}'|{\otimes}{\langle}{\phi}'|{\Pi}_{y}|{\psi}'{\rangle}{\otimes}|{\phi}'{\rangle}
{\langle}{\psi}|{\otimes}{\langle}{\phi}|{\Pi}_{y}|{\psi}{\rangle}{\otimes}|{\phi}{\rangle} \nonumber
\\ &=& 0,
\ee where we have used the Cauchy-Schwarz inequality and Eq. (\ref{cond2}). Applying this to Eq. (\ref{trace2}), we see that the
disjointness of the basis sets implies that all of the expectation values in this sum are equal to zero. This implies that the trace of
${\Pi}_{y}$ is zero, from which we conclude that ${\Pi}_{y}=0$. When this is the case we see, using Eq. (\ref{prob}), that is is
impossible to confirm unambiguously and with non-zero probability that both systems have been prepared in the same state.

We are thus led to consider a two-outcome measurement with
corresponding POVM elements  $\Pi_n$ and $\Pi_?$.  With such a
measurement we will, at most, be able to determine if the states
of both systems are different.  It is important to optimise this
measurement so that the probability of detecting a difference
attains its maximum possible value, which is what we shall now do.

Condition (\ref{cond1}) implies that the support of the operator $\Pi_n$ is a subspace of the antisymmetric subspace. That is,
$$
\Pi_n = \sum_{{\mu}=1}^{D(D-1)/2}e_{\mu} \vert e_{\mu}\rangle \langle e_{\mu}\vert,
$$
for some states $\vert e_{\mu}\rangle$ which form an orthonormal basis for the antisymmetric subspace, and some real, non-negative
coefficients $e_{\mu}$ bounded by $0\le{e_{\mu}}\le 1$. It is a simple matter to show that the maximum probability of any pair of
different states giving rise to an `n' result is attained when all $e_{\mu}$ are equal to 1.  This implies that the optimal POVM
element for detecting differences between the states of the two systems is ${\Pi}_{n}=P_{anti}$, the projector onto the asymmetric
subspace.  From the resolution of the identity and Eq. (\ref{ressym}), we see that its complementary element ${\Pi}_{?}$, which is
responsible for inconclusive results is equal to the projector onto the symmetric subspace, $P_{sym}$. This concludes the matter of
optimisation.

%In addition, because we deal with completely
%unknown pure states, we cannot discriminate between the symmetric states.
%Hence there is no universal test that can identify the states as being identical.
%Our principal ignorance towards the symmetric states forces us
%to rely only on the predictions gained by projecting onto the antisymmetric state. Making the projections onto
%the subspace spanned by the antisymmetric projectors as large as quantum mechanics allows we obtain the optimum
%test for comparison. Due to the base independent character of the measurements projecting onto the
%symmetric and antisymmetric subspace out test is universal, it will perform equally well for any pair of input

To place these findings in context, it is interesting to consider
the related findings recently made by Winter\cite{Winter}.  His
results relate to estimation of the fidelity between two states.
 When specialised to the case of unit fidelity, which corresponds to state comparison, he shows that
 there is no measurement which is symmetrical in its inputs and,
 with unit probability, will reveal whether the states are the
 same or different.  Here, we have not assumed symmetry, nor did we
 require that a conclusive result is obtained with unit
 probability.  Interestingly, Winter showed that the physical
 POVM element which optimally approximates the proposed fidelity
 estimation scheme is $\frac{2}{3}P_{sym}$, which does not appear in the optimal measurement for the criteria we have used.
 This shows that there are at least two, inequivalent approaches
 to the problem of comparing two unknown states.
%states and any measurement basis chosen to project onto.

We can go further in comparing the states of the systems only if
we have {\it {additional information}} about the possible states
that might have been prepared. In particular, if we know that each
system was prepared in a non-degenerate eigenstate of some
Hermitian operator, then we have only to measure the observable
corresponding to the operator on each system and compare the
results. A more subtle and more essentially quantum case involves
a known set of possible non-orthogonal states. As an example, we
consider a pair of two level systems, each prepared in either the
state $\vert\psi_1\rangle$ or the state $\vert\psi_2\rangle$.  In
a suitable orthonormal basis $(\vert +\rangle,\vert -\rangle)$,
and with appropriate adjustment of overall phases, it is always
possible to write such a pair of states in the form \be
\begin{array}{ccl}
\vert\psi_1\rangle &  = & \cos\theta \vert + \rangle +
\sin\theta\vert -
\rangle \\
\vert\psi_2\rangle &  = & \cos\theta \vert + \rangle - \sin\theta\vert - \rangle,
\end{array} \label{def}
\ee for some angle $\theta{\in}[0,\pi/4]$.

For simplicity, we consider only the case in which the states are selected with equal probability. We note that it is not possible to
determine with unit efficiency the state of either system. The best that we can do is either to establish the state with a minimum
probability of error \cite{Helstrom} \be P_{e,min} = \frac{1}{2} [1-\sqrt{1-\vert\langle\psi_1\vert\psi_2\rangle\vert^2}] = \frac{1}{2}
(1-\sin 2\theta ), \ee or to determine the state unambiguously but to accept a minimum probability for the inconclusive result
\cite{Ivanovic,Dieks,Peres} \be P_? = \vert\langle\psi_1\vert\psi_2\rangle \vert = \cos 2\theta . \label{IDP} \ee It is interesting to
note that both forms of quantum state discrimination \cite{Tony} have been demonstrated in experiments with optical polarization
\cite{Huttner,SMB}. We can, of course, determine if the systems have been prepared in the same or different states by measuring each by
means of these optimal single-system strategies. A minimum error single-photon measurement will correctly identify the states as
identical or different if neither or both of the individual measurements is incorrect. This leads to a probability of error in
determining the similarity or difference of the states given by \be P^{comp}_e = \frac{1}{2} \cos^2 (2\theta ) . \ee An attempt based
on unambiguous measurements will fail if either or both of the measurements is inconclusive. This will happen with probability \be
P^{comp}_? = \cos(2\theta ) [2 - \cos (2\theta )] . \ee

Measurement of the individual systems do not necessarily constitute an optimal test of similarity
or difference in the states of the two systems. It is not difficult, however, to derive the
optimal strategies for state comparison. We start by writing the combined state of the two systems
in the form
\begin{eqnarray}
\vert\psi_i\rangle{\otimes}\vert\psi_j\rangle &=&\cos^2\theta
\vert+\rangle{\otimes}\vert+\rangle + (-1)^{i+j}
\sin^2\theta \vert -\rangle{\otimes}\vert -\rangle \nonumber \\
&+& (-1)^{i}\cos\theta \sin\theta[-{\delta}_{ij}(\vert
+\rangle{\otimes}\vert -\rangle + \vert -\rangle{\otimes}\vert
+\rangle  ) \nonumber \\
&+&(1-{\delta}_{ij})(\vert +\rangle{\otimes}\vert -\rangle - \vert -\rangle{\otimes}\vert +\rangle)],
\end{eqnarray}
where $i,j =1,2$. The first three states in this expression are symmetric, while the fourth state is antisymmetric. The antisymmetric
state is unambiguously associated with the states being different and an initial measurement to determine if the combined systems are
in this antisymmetric state will find the answer ``yes" with probability $\frac{1}{2}\sin^2 2\theta$ if the states of the two systems
are different and zero if they are the same. Similarly, the symmetric state $(\vert +\rangle{\otimes}\vert -\rangle + \vert
-\rangle{\otimes}\vert +\rangle )/\sqrt{2}$ is unambiguously associated with the two systems having been prepared in the same state. If
the states were prepared in the same state then this measurement will unambiguously recognize the similarity with probability
$\frac{1}{2}\sin^2 2\theta $. If the combined system is not found in the antisymmetric state or in the symmetric state $(\vert
+\rangle{\otimes}\vert -\rangle + \vert -\rangle{\otimes}\vert +\rangle )/\sqrt{2}$, then it is left in one of the states \be \vert
\Phi_\pm \rangle = \frac{\cos^2\theta\vert +\rangle{\otimes}\vert +\rangle \pm \sin^2\theta \vert -\rangle{\otimes}\vert
-\rangle}{\sqrt{1 - \frac{1}{2}\sin^2 2\theta}}, \ee with the $+ (-)$ sign being associated with the (different) states for the two
qubits. These remaining states lie in a two-dimensional subspace spanned by the symmetric states $\vert +\rangle{\otimes}\vert
+\rangle$ and $\vert -\rangle{\otimes}\vert -\rangle $. Optimal strategies are known for discriminating between such states, either
with minimum error or unambiguously with minimum probability for the inconclusive result \cite{Helstrom,Ivanovic,Dieks,Peres,Tony}.
Combining probability $\frac{1}{2}\sin^2 2\theta$ for finding the combined system in one of the states $(\vert +\rangle{\otimes}\vert
+\rangle \pm \vert -\rangle{\otimes}\vert -\rangle )/\sqrt{2}$ we obtain the minimum probability of error in determining if the states
are the same or different
\begin{eqnarray}
P^{comp}_{e,min}&=&\bigg( 1 -\frac{1}{2} \sin^2 2\theta \bigg) \frac{1}{2} \big[ 1 - \sqrt{1 -
\vert\langle\Phi_+\vert\Phi_-\rangle\vert^2}\big]\nonumber
\\&=& \frac{1}{2} \cos^2 2\theta,
\end{eqnarray}
which is the same as the minimum error probability based on
separate measurements of the single systems (5).   We should note,
however, that the quality of information we receive by the two
methods is quite different. Following the method based on the
symmetry or antisymmetry of the states provides an unambiguous
answer with probability $\frac{1}{2}\sin^2 2\theta$. The price we
pay for this, however, is that we are left to guess if a result
that the states are the same (or different) suggests that the
combined state was $\vert\psi_1\rangle{\otimes}\vert\psi_1\rangle$
or $\vert\psi_2\rangle{\otimes}\vert\psi_2\rangle$ (respectively
$\vert\psi_1\rangle{\otimes}\vert\psi_2\rangle$ or
$\vert\psi_2\rangle{\otimes}\vert\psi_1\rangle$).

To prove that this is indeed the minimum probability of error
attainable using any measurement, we take advantage of the fact
that the state comparison problem can be rephrased as a problem of
distinguishing between two mixed states.  If both systems are
prepared in the same state, with equal probabilities, then the
state of the pair is described by the density operator
\begin{equation}
{\rho}_{s}=\frac{1}{2}\left(|\psi_{1}{\rangle}{\langle}{\psi}_{1}|{\otimes}|\psi_{1}{\rangle}{\langle}{\psi}_{1}|+|\psi_{2}{\rangle}{\langle}{\psi}_{2}|{\otimes}|\psi_{2}{\rangle}{\langle}{\psi}_{2}|\right).
\end{equation}
If, on the other hand, the states of the systems are different,
then the density operator is
\begin{equation}
{\rho}_{a}=\frac{1}{2}\left(|\psi_{1}{\rangle}{\langle}{\psi}_{1}|{\otimes}|\psi_{2}{\rangle}{\langle}{\psi}_{2}|+|\psi_{2}{\rangle}{\langle}{\psi}_{2}|{\otimes}|\psi_{1}{\rangle}{\langle}{\psi}_{1}|\right).
\end{equation}
The problem of determining which density operator applies to the
entire system is clearly equivalent to determining whether or not
both particles are prepared in the same state.  The minimum
probability of error for distinguishing between the symmetric and
asymmetric density operators is given by the Helstrom bound:
\begin{equation}
P_{e,min}=\frac{1}{2}\left(1-\frac{1}{2}\mathrm{Tr}\sqrt{ ({\rho}_{s}-{\rho}_{a})^{2} }\right).
\end{equation}
Substitution of the above expressions for ${\rho}_{s}$ and ${\rho}_{a}$ into the Helstrom bound here gives
$P_{e,min}=(1/2){\cos}^{2}(2\theta)$, confirming the optimality of our measurement.

For unambiguous discrimination we find that it is possible to improve on the probability of inconclusive results based on single-system
measurements. By considering optimal unambiguous discrimination between $\vert\Phi_\pm\rangle$, we find that we can unambiguously
determine if the states of the two systems are the same or different with a minimum probability for inconclusive result given by \be
P^{comp}_{?,min} = \big( 1 - \frac{1}{2}\sin^2 2\theta \big) \vert\langle\Phi_+\vert\Phi_-\rangle\vert = \cos 2\theta. \label{ucb} \ee
This is clearly smaller than the minimum probability of an inconclusive result based on single-system measurements (6). The physical
reason for the better performance of the present strategy compared to the individual unambiguous state discrimination is clear. We can
successfully determine whether the states are the same or different but fail to identify which state label to assign to the individual
systems. This rather obvious lack of information is the reason for the improved performance.  However, it is intriguing that, as we
have shown above, one cannot take advantage of the fact that state comparison is less demanding than individual state discrimination to
improve upon the latter as a technique for minimum-error state comparison.

We will now prove that the bound in Eq. (\ref{ucb}) is optimal.  We do this by showing how an established bound on unambiguous
discrimination leads directly to one for unambiguous comparison, which will turn out to be that given by Eq. (\ref{ucb}). Suppose that
we have an unambiguous comparison machine, where the first system is prepared the {\em known} state $|{\psi}_{1}{\rangle}$ and the
second is in either $|{\psi}_{1}{\rangle}$ or $|{\psi}_{2}{\rangle}$, with equal probability of 1/2.  A machine which can unambiguously
tell us whether the states of both systems are equal or different will then be able to unambiguously discriminate between the possible
states of the second system.

So, let $P_{?1}$ be the probability of inconclusive results in attempting to compare the states of the two systems when the first
system is in the state $|{\psi}_{1}{\rangle}$.  Whenever the unambiguous comparison attempt is successful, it leads to successful
unambiguous discrimination between the possible states of the second system.  It follows that any universal bound on the probability of
unambiguous discrimination between two pure states must apply to the particular setup considered here.  We have seen that
${\cos}2{\theta}$ is a lower bound on the minimum probability of inconclusive results when attempting to unambiguously discriminate
$|{\psi}_{1}{\rangle}$ and $|{\psi}_{2}{\rangle}$ in Eq. ({\ref{IDP}}).  This gives \be P_{?1}{\geq}{\cos}2{\theta}. \ee We can repeat
the argument when the first system is known to be prepared in the state $|{\psi}_{2}{\rangle}$ instead.  Again, under these
circumstances, unambiguous comparison of the states of both systems enables us to unambiguously discriminate between the possible
states of the second system.  Letting $P_{?2}$ be the probability of unambiguous comparison when the first system is in the state
$|{\psi}_{2}{\rangle}$ and the second is in either $|{\psi}_{1}{\rangle}$ or $|{\psi}_{2}{\rangle}$, with equal probability of 1/2, the
optimality of the bound in Eq. ({\ref{IDP}) implies that \be P_{?2}{\geq}{\cos}2{\theta}. \ee Suppose now that the state of the first
system is also $|{\psi}_{1}{\rangle}$ or $|{\psi}_{2}{\rangle}$, with equal probability of 1/2 so that the four possible product states
of the composite system have equal probability.   The probability of inconclusive results in an  unambiguous comparison attempt is then
$P_{?}^{comp}=(P_{?1}+P_{?2})/2$.  Making us of this, and combining it with the above bounds on $P_{?1}$ and $P_{?2}$, we finally
obtain the following lower bound on $P_{?}^{comp}$: \be P_{?}^{comp}=\frac{P_{?1}+P_{?2}}{2}{\geq}{\cos}2{\theta}, \ee which is to say
that the minimum probability of inconclusive results in unambiguous discrimination between two states is a lower bound on that of those
obtained in unambiguous comparison. Having demonstrated in Eq. ({\ref{ucb}}) that this bound is attainable, it is clearly seen to be
optimal.

Quantum limited state comparison can be realized, at least in
part, with current experimental techniques. In particular, two
photons incident on a $50/50$ beam splitter will exit in the same
direction if they are in a symmetric polarization state but will
leave in different directions if they are in the antisymmetric
state of polarization \cite{Weihs}. This effect can be used to
test if two photons were prepared in the same state of
polarization. For two photons, each prepared in one of two known
quantum states, we can determine the state to be the same or
different with minimum probability of error by simply determining
the state of each photon with minimum error \cite{SMB}. Performing
the minimum error state comparison based on the symmetry, however,
requires us to make a full Bell-state measurement.  When our
experimental techniques are restricted to linear optics, this is
not possible when the Bell states are encoded only in optical
polarization\cite{Kalle}.  However, Kwiat and
Weinfurter\cite{Kwiat} showed that, if the Bell states are encoded
in a higher dimensional Hilbert space where the photons are also
momentum-entangled, then the full Bell measurement can be carried
out using linear optics.  In the near future, it may be possible
to implement the full Bell measurement and optimal unambiguous
state comparison for two possible states using systems other than
optical polarization. Possible candidates include trapped ions and
NMR with which manipulation of entanglement has been shown to be
possible (see \cite{Wineland,NC} and references therein.)

We should emphasise that if two systems are prepared in the same
state, the composite state is, in general, confined to the
symmetric subspace only if it is pure.   The situation is
different for mixed states.  For example, two unpolarized
spin-half particles may each be described by the same density
operator $\frac{1}{2}1$. The combined density operator
$\frac{1}{4} 1{\otimes}1$ has a nonzero overlap with the
antisymmetric state even though the mixed states of the two
particles are identical. This means that comparison of systems
prepared in entangled states will require access to all the
entangled subsystems if a state comparison is to be made. The
comparison of entangled states leads us already beyond the scope
of problems discussed in the present paper. Here we have focused
mainly on the problem of two qubits. The comparison of at least
two entangled states can be fully understood only within a more
general approach allowing for systems to have more levels and (or)
treating the comparison of several copies. A detailed treatment of
this subject will be given elsewhere.

We conclude by suggesting some potential applications of quantum state comparison. In cryptography \cite{crypt} it may perhaps be
beneficial for an eavesdropper to compare successive photons rather than to measure individual photons and any secure protocol must
guard against such an attack. The usefulness of state comparison in relation to quantum fingerprinting protocols has been discussed by
Buhrman {\em et al}\cite{finger}. Comparison between two quantum devices, such as quantum computers, could provide a method to detect
errors in one of them. Finally, at the limits of metrology, small effects might be monitored by quantum state comparison with a highly
reliable reference system.

\noindent {\it Acknowledgements}
We thank Erika Andersson for helpful comments and M. Hillery and J. Bergou for discussions.
This work was supported by the UK Engineering and Physical Sciences Research Council, the Royal Society of
Edinburgh, the Scottish Executive Education and Lifelong Learning Department.
The financial support by the M\v SM 210000018
(Czech Republic), GA\v CR 202/01/0318 and EU IST-1999-13021 for I.J. is gratefully acknowledged.

\end{document}